\documentclass[twocolumn]{article}
\usepackage[nointlimits]{amsmath}
\usepackage{amsfonts}
\usepackage{amssymb}
\usepackage{amsmath}
\usepackage{amsthm}
\usepackage{latexsym}
\usepackage{graphicx}
\usepackage{layout}
\usepackage[english]{babel}

\newtheorem{lemma1}     {Lemma}[section]
\newtheorem{teorema1}   [lemma1]{Theorem}
\newtheorem{prop1}      [lemma1]{Proposition}
\newtheorem{coroll1}    [lemma1]{Corollary}
\newtheorem{cong1}      [lemma1]{Conjecture}
\newtheorem{remark1}    [lemma1]{Remark}
\newtheorem{defin1}     [lemma1]{Definition}

\newenvironment{Theorem}[1][]
        {\begin{teorema1}[#1]\begin{samepage}}{\end{samepage}\end{teorema1}}
\newenvironment{Proposition}[1][]
        {\begin{prop1}[#1]\begin{samepage}}{\end{samepage}\end{prop1}}

\newenvironment{Remark}[1][]
        {\begin{remark1}[#1]\begin{samepage}}{\end{samepage}\end{remark1}}
\newenvironment{Definition}[1][]
        {\begin{defin1}[#1]\begin{samepage}}{\end{samepage}\end{defin1}}

\newcommand{\nada}[1]   {}

\newcommand{\wT}         {\widetilde T}

\newcommand{\om}        {\omega}

\renewcommand{\d}         {{\rm d}}
\newcommand{\R}         {\ensuremath{\mathbb R}}

\newcommand{\uu}         {{\bf u}}
\newcommand{\ff}         {{\bf f}}
\newcommand{\phiphi}    {\mathbf{v}}

\newcommand{\tensoremetrico}    {\eta}

\newcommand{\res}      {\mathop{\hbox{\vrule height 7pt width .5pt 
                        depth 0pt \vrule height .5pt width 6pt depth 0pt}}\nolimits}
\newcommand{\minsurf}        {M} 
\newcommand{\timelikeminsurf}        {\Sigma} 
\newcommand{\lorentztransf}    {L}
\newcommand{\minkproj}    {P}
\newcommand{\Rn}        {\ensuremath{\mathbb R^n}}

\renewcommand{\H}       {\mathcal{H}}

\newcommand{\eps}       {\epsilon}

\newcommand{\grad}      {\nabla}

\begin{document}
\title{\textbf{Time-like 
minimal submanifolds as singular limits of nonlinear wave equations}}

\author{Giovanni Bellettini
\thanks{Dipartimento di Matematica, Universit\`a di Roma ``Tor
  Vergata'', Via della Ricerca Scientifica, 00133 Roma, Italy,  
and Laboratori Nazionali di Frascati dell'INFN,
via E. Fermi, 40, 00044 Frascati (Roma), Italy,
  \texttt{Giovanni.Bellettini@lnf.infn.it}} 
\and Matteo Novaga
\thanks{Dipartimento di Matematica, Universit\`a di Padova, 
	Via Trieste 63, 35121 Padova, Italy, 
  \texttt{novaga@math.unipd.it}} 
\and Giandomenico Orlandi 
\thanks{Dipartimento di Informatica, Universit\`a di Verona,
  Ca' Vignal 2, strada le Grazie 15, 37134 Verona, Italy,
  \texttt{giandomenico.orlandi@univr.it}}}

\date{}
\maketitle

\begin{abstract}
\noindent 
We consider the sharp interface limit $\eps \to 0^+$ 
of the semilinear wave equation
$\Box \uu + \nabla W(\uu)/\eps^2=0$ in $\R^{1+n}$, 
where 
$\uu$ takes
values in $\R^k$, $k=1,2$,
and $W$ is a double-well potential if $k=1$ and vanishes 
on the unit circle and is positive elsewhere if $k=2$.
For fixed $\eps >0$ we find some special solutions, constructed around
minimal surfaces in $\Rn$. 
In the general case, under some additional assumptions, we show that
the solutions converge to a Radon 
measure supported on 
a time-like $k$-codimensional 
minimal submanifold of the Minkowski 
space-time. This result holds also after the appearence 
of singularities, and 
 enforces the observation made by J. Neu that this 
semilinear equation can be regarded as an approximation of the Born-Infeld equation.
\end{abstract}

\section{Introduction}
In this paper we consider the following system of semilinear 
hyperbolic equations 
\begin{equation}\label{eqhyp}
\Box \uu + \frac{1}{\eps^2} \nabla W(\uu)= 0\,,
\end{equation}
for 
$$
\uu:  \R\times\R^{n}  \to \R^k,
\qquad n \geq 1, \ k =1,2,
$$
where $\Box \uu = \uu_{tt}-\Delta \uu =  
\partial_{x^0 x^0} \uu - \partial_{x^i x^i} \uu$
is the wave operator in $\R^{1+n}$ with coordinates $x^0=t, x^1,\dots,x^n$,
$\eps>0$ is a small parameter,
and $W(\uu)=\widetilde W(|\uu|)$, 
where $\widetilde W:\R\to \R^+$ is a double-well potential. 
Equation \eqref{eqhyp} is a Lorentz invariant field equation, 
governing the dynamics of topological defects
such as vortices \cite{neuuno}; it is also strictly
related to time-like lorentzian minimal submanifolds of codimension $k$
in Minkowski $(1+n)$-dimensional space-time \cite{neu}. 
We refer to \cite{ShSt} for a discussion on the existence
of local and global solutions to \eqref{eqhyp}. 
The elliptic/parabolic analog of \eqref{eqhyp} is called
the Ginzburg-Landau equation, and
has been recently investigated
by many authors in connection with euclidean minimal surfaces and mean curvature 
flow in codimension $k$ (see for instance 
\cite{BeGiSm:06} and references therein).
Here we are interested in the asymptotic limit
as $\eps \to 0^+$ of solutions $\uu_\eps$ to \eqref{eqhyp}.
The case $k=1$ will be referred to as the scalar case, since
\eqref{eqhyp} reduces to a single equation, and solutions 
will be denoted by $u_\eps$; note that 
in this case, the vacuum states $\pm 1$ 
are stable solutions. 

For $n=3$ and $k=1$, the 
asymptotic limit as $\eps\to 0^+$ of $u_\eps$
has been formally computed by Neu in \cite{neu}, 
using suitable asymptotic expansions. 
The author shows that there are solutions which take
the constant values $\pm 1$ out of a transition layer 
of thickness $\eps$, provided such a layer is suitably 
close to a one-codimensional
time-like lorentzian minimal surface $\Sigma$.
The one-codimensional
time-like minimal surface equation 
can be described as follows: 
the points 
$(x^0,x^1,\cdots,x^n)$ on each time-slice
$\Sigma(t) := \Sigma \cap \{ x^0 = t\}$
of $\Sigma$  must satisfy the equation
\begin{equation}\label{eveq}
A = (1-V^2) \kappa 
\end{equation}
in normal direction, where $A$, $V$ and $\kappa$ 
are respectively the acceleration, the velocity
and the euclidean mean curvature  of
$\Sigma(t)$ at $(x^0,x^1,\cdots,x^n)$.
We point out that Eq. \eqref{eveq} is the Euler-Lagrange equation
of the $n$-dimensional area induced by the Minkowski metric, given by
\[
\mathcal A(\Sigma) = \int_\Sigma \sqrt{-\nu_0^2 + |\hat\nu|^2}~ d\mathcal H^{n},
\]
where $\nu=(\nu_0,\hat\nu)$ is a unit euclidean normal to $\Sigma$, 
and ${\mathcal H}^n$ is the $n$-dimensional euclidean Hausdorff 
measure. 
We refer to \cite{BoIn:34}, \cite{Li:04},
\cite{Ho:94} for the analysis of various aspects of Eq. \eqref{eveq}. 
Interestingly, Neu \cite{neu} showed also that, due to 
possible oscillations on a small scale on the initial interface,
which are not dissipated in time,   
solutions to \eqref{eqhyp} may not converge 
to a solution of \eqref{eveq}, 
as the oscillation scale tends to zero. 

In the first part of the present paper we compute some explicit 
selfsimilar solutions of \eqref{eqhyp}. In particular, 
we show that, given any 
euclidean nondegenerate minimal hypersurface $\minsurf$ in $\R^n$,  
there exists a solution to \eqref{eqhyp} traveling around
$\minsurf$ (see Propositions 
\ref{teoTW} and \ref{prodo}).

In the second part of the paper we adapt to the hyperbolic setting
the parabolic strategy followed   
in \cite{AmSo:98}.
Given a solution $\uu_\eps$ to \eqref{eqhyp} let 
\begin{equation*}
\ell_\eps(\uu_\eps) :=
c_k(\eps)\left( \frac{-|{\uu_\eps}_t|^2
+ \vert\nabla \uu_\eps\vert^2}{2} + \frac{W(\uu_\eps)}{\eps^2}\right)
\end{equation*}
be the rescaled lagrangian integrand, where 
\[
c_k(\eps) := \left\{ 
\begin{array}{ll}
\eps & {\rm if}\ k=1,
\\
\frac{1}{|\log\eps|} & {\rm if}\ k=2.
\end{array}\right.
\]
In our main 
result (Theorem \ref{teomain}) 
 we show that, under some technical assumptions, 
$\ell_\eps(\uu_\eps)$ 
concentrates on a $k$-codimensional
set $\Gamma$, as $\eps\to 0^+$.
Moreover, $\Gamma$ is a time-like lorentzian minimal submanifold 
whenever it is smooth. In order to prove this result we suitably extend 
the notion of rectifiable varifold to the lorentzian setting, and 
prove that the stress-energy tensor of the solutions of \eqref{eqhyp} converges  
to a stationary lorentzian varifold, as $\eps \to 0^+$.
The proof of Theorem \ref{teomain} naturally leads to Definition \ref{defva},
which generalizes the concept of minimality, with respect to the Minkowski area, 
to nonsmooth $k$-codimensional submanifolds.
A weak notion of lorentzian minimal submanifold 
(a lorentzian stationary varifold in our case) 
seems here to be unavoidable, in view of the presence
of singularities.

Finally, we conclude the paper by discussing the validity of our assumptions 
in relation to the example of Neu \cite{neu}.

\subsection{Notation}
 
Throughout the paper bold letters will
refer to the case $k=2$.
The greek indices $\alpha, \beta, \gamma, \delta$ run from $0$ to $n$, while
the roman indices $i,j$ run from $1$ to $n$; we adopt 
the Einstein summation convention over repeated indices.

We let $\tensoremetrico^{-1} ={\rm diag}(-1,1,\dots,1)$ be the inverse
Minkowski metric tensor with contravariant components $\eta^{\alpha\beta}$;
$\eta_{\alpha\beta}$ are the covariant components of the 
Minkowski metric tensor $\eta$.

Given $\xi= (\xi_0, \hat \xi) \in \R\times \R^{n}$
we set $\vert \hat \xi\vert^2 := \eta^{ij} \hat \xi_i \hat \xi_j$,
\begin{equation*}
\langle \xi,\xi \rangle_m 
:= -(\xi_0)^2 + \vert \hat
\xi\vert^2 = \tensoremetrico^{\alpha\beta}\xi_\alpha\xi_\beta,
\end{equation*}
and if $\langle \xi,\xi \rangle_m \neq 0$ we set
$|\xi|_m := \frac{\langle \xi,\xi \rangle_m}{\vert\langle \xi,\xi 
\rangle_m\vert^{1/2}}$.
We say that $\xi$ is space-like 
(resp. time-like) if $\langle \xi,\xi \rangle_m >0$ 
(resp. $\langle \xi,\xi \rangle_m<0$).
Given a $(1,1)$-tensor $A$, we 
say that $A$ is space-like 
(resp. time-like) if $A\xi$ is space-like 
(resp. time-like) for all $\xi\in\R \times \R^n  \setminus \{(0,0)\}$.
 
$\nabla$ (resp. $\overline \nabla$) 
indicates the euclidean gradient in $\Rn$ 
(resp. in $\R^{1+n}$); for a smooth function $g : \R^{1+n}
\to \R$ we set $\nabla_m g := (-g_t, \nabla g) = \tensoremetrico^{\alpha\beta}
\frac{\partial g}{\partial x^\beta} = \eta^{-1} \overline \nabla g$.

$\mathcal H^h$ denotes the $h$-dimensional euclidean area (i.e. the 
Hausdorff measure)
either in $\Rn$ or in  $\R^{1+n}$ for $h \in \{0,\dots,n\}$;
 $\res$ is the symbol of restriction of measures 
and $\rightharpoonup$ denotes the weak convergence of Radon measures. 
If $\mu$ is a measure absolutely continuous with respect to $\lambda$,
we write $\mu << \lambda$, and
we denote by $d\mu/d\lambda$ the Radon-Nikodym derivative of 
$\mu$ with respect to $\lambda$.

We recall that a smooth $k$-codimensional submanifold $\minsurf$ of $\R^n$ 
without boundary is 
said minimal
if $\minsurf$ has vanishing mean curvature. A minimal submanifold $\minsurf
\subset \Rn$ 
is said nondegenerate 
if the second variation of its $(n-k)$-dimensional area,
represented by the associated Jacobi operator, is injective.

\section{Selfsimilar solutions}\label{sectrav}

Unless otherwise specified,
in what follows we take $W(\uu) = \frac{1}{4} (1-|\uu|^2)^2$ if $n \leq 4$, and
if $n>4$ we suppose $W$ to be a function of $\vert \uu\vert$
with the proper growth at infinity
in order problem \eqref{eqhyp} to be well-posed \cite{ShSt}.

We let  
\begin{equation*}
e_\eps(\uu_\eps) := 
c_k(\eps)\left( \frac{|{\uu_\eps}_t|^2
+ \vert\nabla \uu_\eps\vert^2}{2} + \frac{W(\uu_\eps)}{\eps^2}\right)
\end{equation*}
be the rescaled energy integrand of a solution $\uu_\eps$ of \eqref{eqhyp}.
By $\vert {\uu_\eps}_t\vert^2$ (resp.
$\vert \nabla \uu_\eps\vert^2$)  we mean the square euclidean norm of
${\uu_\eps}_t \in \R^k$ (resp. of
 $\nabla \uu_\eps$,
i.e.,
the sum of the squares of the 
elements of the matrix $\nabla \uu_\eps$).

We notice that 
the following quantity is
conserved for any $t \geq 0$: 
\begin{equation}\label{eqcon}
\int_{\R^n} e_\eps(\uu_\eps(t,x))~dx 
= \int_{\R^n} e_\eps(\uu_\eps(0,x))~dx,
\end{equation}
assuming the proper growth conditions on the right hand side.

\subsection{Traveling waves}

Let $k=1,2$. 
We construct solutions of \eqref{eqhyp}, which are traveling waves 
along a prescribed direction $\nu\in\R^n$, $|\nu|=1$. 
Up to a rotation of $\R^n$, we can assume $\nu=(0,\ldots,0,1)$. 
Letting $x=(y,z)\in \R^n=\R^{n-1}\times \R$,
we look for traveling wave solutions of \eqref{eqhyp} of the form 
\begin{equation}\label{uphi}
\uu_\eps(t,x)=\phiphi(y,z-vt),
\end{equation}
for some $v\in (-1,1)$ and a suitable map $\phiphi: \Rn \to \R^k$. 
Then \eqref{eqhyp} becomes 
\begin{equation}\label{eqtrav}
-\Delta_y \phiphi - (1-v^2) \phiphi_{zz}+ \frac{1}{\eps^2} \nabla W(\phiphi) 
= 0,
\end{equation}
where $\Delta_y$ is the Laplacian in $\R^{n-1}$ with respect to the 
$y = (y^1,\dots,y^{n-1})$-coordinates.
Let 
\begin{equation}\label{wphi}
\ff(y,z) := \phiphi(y,\sqrt{1-v^2}z).
\end{equation}
Then $\ff$ satisfies the elliptic Ginzburg-Landau system
\begin{equation}\label{eqGL}
-\Delta \ff + \frac{1}{\eps^2} \nabla W(\ff) = 0. 
\end{equation}
Hence
traveling wave solutions of \eqref{eqhyp}, with $v\in (-1,1)$,
correspond to solutions of the elliptic  system \eqref{eqGL}.

We recall the following result \cite{Pacard}.

\begin{Theorem}\label{probrendle}
For any smooth nondegenerate embedded minimal 
submanifold $\minsurf\subset\R^n$ of codimension $1$
without boundary, 
there exist solutions $f_\eps$ of
\eqref{eqGL} such that 
\[
\eps\, \left( \frac{\vert\nabla f_\eps\vert^2}{2} 
+ \frac{W(f_\eps)}{\eps^2}\right) 
 \rightharpoonup \sigma\, \mathcal H^{n-1}\res \minsurf
\]
as $\eps\to 0^+$, where $\sigma =\sigma(W,n)$ is a positive constant
independent of $\minsurf$.
\end{Theorem}
As a consequence our first result is the 
existence of traveling waves close to any 
nondegenerate minimal hypersurface of $\R^n$.
\begin{Proposition}\label{teoTW}
Let $k=1$. Let $\minsurf \subset \R^n$ be 
a smooth nondegenerate 
embedded minimal submanifold of codimension $1$ without 
boundary, and let $v \in (-1,1)$. Define
\begin{align*}
\timelikeminsurf := & 
\Big\{ \left(t,y, \sqrt{1-v^2}z + vt\right) \in \R \times \R^{n-1}
\times \R :
\\ 
& \ \ \ (y,z)\in \minsurf\Big\}.
\end{align*}
Then there exist traveling wave solutions 
$u_\eps:\R^{1+n}\to \R$ of \eqref{eqhyp} of the form \eqref{uphi}, 
such that 
\begin{eqnarray}\label{eell}
\ell_\eps(u_\eps) \rightharpoonup \sigma\,\mu\res\timelikeminsurf
\qquad {\rm}\ \eps\to 0^+,
\end{eqnarray}
where the measure $\mu$ 
is the $n$-dimensional area induced by the Minkowski metric.
\end{Proposition}

\begin{proof}
Set $\gamma := (1-v^2)^{-1/2}$. 
If $f_\eps$ are as in Theorem \ref{probrendle}, we define 
$u_\eps(t,x) := f_\eps(y,\gamma(z-vt))$.
Then $\ell_\eps(u_\eps) = 
\eps\, \left( \frac{\vert\nabla f_\eps\vert^2}{2} 
+ \frac{W(f_\eps)}{\eps^2}\right)$, hence if $\varphi 
\in C_c(\R^{1+n})$,
\begin{align}
\nonumber
& \int_\R \int_{\Rn} \ell_\eps(u_\eps) \varphi ~dxdt 
\\
\label{grdd}
\\
= & 
\int_\R \int_{\Rn} 
\eps\, \left( \frac{\vert\nabla f_\eps\vert^2}{2} 
+ \frac{W(f_\eps)}{\eps^2}\right)\varphi
~dxdt 
\nonumber
\end{align}
where the integrand in parentheses is evaluated at
$(y, \gamma(z-vt))$. Therefore, making the change of variables 
$(t',y',z')= L(t,y,z)$,  
where $L$ is the Lorentz transformation given by 
\[
L(t,y,z):=\left( \gamma(t-vz),y,\gamma(z-vt) \right),
\]
we have that \eqref{grdd} equals
\begin{align*}
& 
\int_\R 
\int_{\Rn} 
\eps\, \left( \frac{\vert\nabla f_\eps\vert^2}{2} 
+ \frac{W(f_\eps)}{\eps^2}\right) \varphi
~dx' dt'
\\
& \to 
\sigma  
\int_\R \int_{\minsurf} \varphi~d \mathcal H^{n-1}(x')\, dt'
 = \sigma \int_\timelikeminsurf 
 \varphi ~d\mu,
\end{align*}
where $\mu$ is the image of $\mathcal H^{n}\res \R\times M$, 
through the map $L^{-1}$.
\end{proof}

\begin{Remark}\rm 
The hypersurface $\timelikeminsurf$ in Proposition \ref{teoTW} 
is a time-like lorentzian minimal hypersurface. Indeed, 
let $d: \Rn \to  \R$ be the signed (euclidean) distance function 
from $\minsurf$, negative in the interior of $\minsurf$ (note that
$\minsurf$ is the boundary of a smooth open subset of $\Rn$), so that 
$\minsurf = \{(y,z) \in \Rn : d(y,z)=0\}$, $\vert \nabla d\vert^2=1$
in a neighbourhood of $\minsurf$,  and $\Delta d =0$ on $\minsurf$. 
Define $g : \R^{1+n} \to \R$ as $g(t,x) := d(y, \gamma(z-vt))$, 
$x = (y,z)$. 
Observe that $\timelikeminsurf 
 = \{g=0\}$, so that the minkowskian mean
curvature of $\timelikeminsurf$ is given by the euclidean divergence 
in $\R^{1+n}$ 
of $\nabla_m g/ \vert \nabla_m g \vert_m$, namely by
$$
\Big(
\frac{-g_t}{\sqrt{- (g_t)^2 + \vert \nabla g\vert^2}} 
\Big)_t + \Big(
\frac{g_{x^i}}{ \sqrt{- (g_t)^2 + \vert \nabla g\vert^2}} 
\Big)_{x^i} 
$$
evaluated on $\timelikeminsurf$. The equality $\vert \nabla d\vert^2=1$ 
implies $\sqrt{- (g_t)^2 + \vert \nabla g\vert^2}=1$ in a neighbourhood
of $\timelikeminsurf$. Therefore we only have to check that 
\begin{equation}\label{cisi}
-
g_{tt} + 
g_{x^i x^i} =0 \qquad {\rm on}~ \timelikeminsurf,
\end{equation}
which is verified because 
$-
g_{tt} + 
g_{x^i x^i}$ on $\timelikeminsurf$ coincides 
with $\Delta d$ on $\minsurf$.
\end{Remark}

Note  
that $\ell_\eps(u_\eps)$ concentrates
  on $\timelikeminsurf$
in the limit $\eps\to 0^+$;
the same happens for 
$e_\eps(u_\eps)$, 
since 
$e_\eps(u_\eps)$, 
and $\ell_\eps(u_\eps)$ in Proposition \ref{teoTW}
are mutually absolutely continuous.

\subsection{Rotating waves}

In this section we let $W(\uu)= (1-\vert \uu\vert^2)^2/4$, 
$\widetilde W:\R\to \R$ be defined as 
$\widetilde W(s):=(1-s^2)^2/4$,
and let $k=2$; we identify the target space $\R^2$ with the complex plane.
We look for solutions of \eqref{eqhyp} of the form 
\begin{equation}\label{eqrot}
\uu_\eps(t,x) = \rho(x)e^{i\om t}, \qquad \rho:\,\R^n\to\R,
\end{equation}
for some $\om\in\R$. Substituting \eqref{eqrot} into \eqref{eqhyp}, 
we get that $\rho$ must satisfy
\begin{equation}\label{eqsol}
-\Delta \rho -\om^2 \rho + \frac{1}{\eps^2} \widetilde W'(\rho) = 0.
\end{equation}
This scalar equation can be rewritten as
\begin{equation}\label{aceps}
-\Delta \rho + \frac{1}{\eps^2} \widetilde W'_\eps(\rho) = 0,
\end{equation}
where
\begin{eqnarray*}
\widetilde W_\eps(\rho) &:=& \frac{(1 + \eps^2\omega^2 -\rho^2)^2}{4}
\\
&=& (1 + \eps^2\omega^2)^2\,\widetilde W\left(\frac{\rho}{\sqrt{1 + \eps^2\omega^2}}\right).
\end{eqnarray*}

Therefore 
\eqref{aceps} reduces to \eqref{eqGL}
with $k=1$ and $W$ replaced by $\widetilde W$,
after the change of variables 
\[
f(x)=  \frac{\rho \left(\frac{x}{\sqrt{1 + \eps^2\omega^2}}\right)}
{\sqrt{1 + \eps^2\omega^2}},
\]
and we can still apply Theorem \ref{probrendle}. 
In particular, we get the following 

\begin{Proposition}\label{prodo}
Let $\minsurf\subset\R^n$  be 
 a smooth nondegenerate embedded minimal submanifold 
of codimension $1$ without boundary, and let $\omega\in\R$. 
Define 
$$
\timelikeminsurf:= \R\times \minsurf.
$$
Then 
there exist solutions $\uu_\eps:\R^{1+n}\to \R^2$ 
of \eqref{eqhyp} of the form \eqref{eqrot}, 
such that 
\[
\eps\left( \frac{-\vert {\uu_\eps}_t\vert^2
+ \vert\nabla \uu_\eps\vert^2}{2}\right) + \frac{W(\uu_\eps)}{\eps}
\rightharpoonup \sigma\,\mu\res \Sigma  
\]
as $\eps\to 0^+$, where the measure $\mu$ 
is the $n$-dimensional area induced by the Minkowski metric.
\end{Proposition}

\begin{proof}
If $\varphi \in C^\infty_c(\R^{1+n})$ we have 
\begin{align*}
& \int_\R
\int_{\Rn} 
\frac{\eps}{c_2(\eps)} \ell_\eps(\uu_\eps)
\varphi
~dxdt 
\\
=&  \int_\R \int_{\Rn} 
\Big[\eps \frac{\vert \nabla \rho\vert^2}{2}
\\
&  + \frac{1}{\eps}\left(
\widetilde W(\rho) + \eps^2 \frac{\rho^2 \omega^2}{2}\right) \Big]
\varphi~dx dt
\\
\to & \ \sigma\int_\R \int_{\minsurf} \varphi ~d\mathcal H^{n-1} dt
= \sigma\,\int_\Sigma \varphi ~d\mu.
\end{align*}
\end{proof}

Note that in Proposition \ref{prodo} $\frac{\eps}{c_2(\eps)}
\ell_\eps(\uu_\eps)$ concentrates on 
the lorentzian 
minimal  submanifold $\timelikeminsurf$ of codimension $1$, even if $k=2$.

\section{Convergence as $\eps\to 0^+$}\label{secconv}

We are interested in passing to the limit in \eqref{eqhyp}, 
as $\eps\to 0^+$. As already mentioned in the 
introduction, a formal limit has been performed in 
\cite{neu} when $k=1$. Rigorous asymptotic results for well prepared initial data
have been recently announced in \cite{Je:08}. 

{}From now on we shall assume that 
there exists a constant $C>0$ such that 
\begin{equation}\label{ipoini}
\sup_{\eps \in (0,1)
}\int_{\R^n}e_\eps(\uu_\eps(0,x))~dx\le C.
\end{equation}

\subsection{Assumptions on $\ell$ and $e$}

Under assumption 
\eqref{ipoini},
from \eqref{eqcon} it follows 
that the measures $e_\eps(\uu_\eps)\ dtdx$ 
converge, up to a (not relabelled) subsequence as $\eps\to 0^+$,
to a Radon measure $e$ in $\R^{1+n}$.
%
%
Since $\vert \ell_\eps(\uu_\eps)\vert$ and 
$c_k(\eps)W(\uu_\eps)/\eps^2$ 
are both bounded by $e_\eps(\uu_\eps)$,
they also weakly converge,
up to a subsequence, to two measures $\ell$ and $w$ respectively, 
\begin{align}
\label{deflw}
& \ell_\eps(\uu_\eps)~dtdx 
\rightharpoonup \ell, 
\\
& c_k(\eps)W(\uu_\eps)/\eps^2~dtdx 
 \rightharpoonup w, 
\label{deflwbis}
\end{align}
and $\ell$ and $w$ are absolutely continuous with respect to $e$, 
with density less than or equal to $1$. In the following,
we shall assume that
\begin{itemize}
\item[(A1)] $e$ is absolutely continuous with respect to $\ell$.
\end{itemize}
Assumption (A1) is a weak way to say that $\vert {\uu_\eps}_t\vert^2$
is controlled by $\vert \grad \uu_\eps\vert^2$, or equivalently 
that the tensor $\overline\grad\uu_\eps$, suitably normalized, becomes space-like 
in the limit $\eps \to 0^+$. Such an assumption essentially implies that the 
set $\Gamma$ defined in \eqref{defGamma} below is time-like. 

Following \cite{AmSo:98}, we shall assume also that
\begin{itemize} 
\item[(A2)]
for $\ell$-almost every $(t,x)$ it holds
\begin{equation}\label{req:A2}
0< \lim_{\rho\to 0^+}
\,\frac{\ell(B_\rho(t,x))}{\rho^{n+1-k}}<+\infty, 
\end{equation}
\end{itemize}
where $B_\rho(t,x)$ denotes the euclidean ball of radius $\rho$ centered at $(t,x)$. 
Recalling Preiss' Theorem \cite{DL:06},
from (A2) it follows that the support of the measures $e$ and $\ell$ 
\begin{equation}\label{defGamma}
\Gamma:= {\rm spt}(\ell) = {\rm spt}(e)
\end{equation}
is a rectifiable set of dimension $n+1-k$, and 
\[
\ell = f \, \mathcal H^{n+1-k}\res \Gamma
\]
in the sense of measures, for some Borel measurable function $f>0$.

Assumption (A2) also ensures that 
the lagrangian integrands 
$\ell_\eps(\uu_\eps)$ concentrate on a time-like rectifiable set $\Gamma$ of 
codimension $k$ in the limit
$\eps \to 0^+$. Hence $\Gamma$ has the correct codimension, but
is not necessarily smooth everywhere. We observe  at this point that, 
in general, time-like lorentzian minimal submanifolds are 
singular,  and that the 
density $f$ defined above is typically $0$
at the singular points of $\Gamma$; for instance
the limit in \eqref{req:A2} vanishes if $(t,x)$ is the vertex of 
half a light-cone. 
 Note carefully that 
we are not excluding the presence of zero density points on $\Gamma$,
since (A2) is required to be valid only for $\ell$-almost
all points, and not for all points of $\Gamma$. 
Differently with respect to the parabolic
case considered in \cite{AmSo:98}, 
we cannot expect here a uniform lower density bound
(where the zero on the left hand side of \eqref{req:A2} 
would be
substituted by an absolute positive constant).

\subsection{Rectifiable lorentzian varifolds}\label{seclorvar}
A matrix $\minkproj$ represents a lorentzian orthogonal 
projection on a time-like subspace of codimension $k$ of $\R^{1+n}$ if 
there exists a 
Lorentz transformation $\lorentztransf$ such that
\[
\lorentztransf^{-1}\, \minkproj\, \lorentztransf = 
\begin{cases}
{\rm diag}(1,0,1,\ldots,1) \qquad {\rm if}~ k=1,
\\
{\rm diag}(1,0,0,1,\ldots,1) \quad {\rm if}~ k=2.
\end{cases}
\]
The pair of Radon measures $V=(\mu^{}_V,\delta_\minkproj)$ is called 
\emph{rectifiable lorentzian varifold} (without boundary) 
of codimension $k$ if ${\rm spt}(\mu^{}_V)
\subset \R^{1+n}$ is an $(n+1-k)$-rectifiable set
whose tangent space is time-like $\mathcal H^{n+1-k}$-almost everywhere, 
and $\delta_\minkproj$ is the Dirac delta concentrated in 
$\minkproj$, where 
$\minkproj$ 
is the lorentzian orthogonal projection onto 
the tangent space to ${\rm spt}(\mu_V)$. Notice that, when $k=1$ and 
${\rm spt}(\mu_V)$ is smooth and time-like, 
the orthogonal lorentzian projection $P$ can be written as 
\[
P = {\rm Id} - \eta^{-1}\nu_m \otimes \nu_m,
\]
where $\nu$ is a normal (co)vector to ${\rm spt}(\mu_V)$, and $\nu_m := \nu/|\nu|_m$.

\begin{Definition}\label{defva}
We say that the rectifiable lorentzian varifold $V = (\mu_V, \delta_P)$ is stationary if 
\begin{equation}\label{eqsta}
\int_{\R^{1+n}} {\rm tr}
\left(\minkproj  ~\overline \nabla {\bf X}\right)\,d\mu^{}_V = 0
\end{equation}
for all vector fields ${\bf X}\in (C^1_c(\R^{1+n}))^{n+1}$.
\end{Definition}
Notice that \eqref{eqsta} is equivalent to require that the generalized 
varifold $(\mu^{}_V,\delta_{\minkproj})$ 
(in the sense of \cite[Def. 3.4]{AmSo:98}) is stationary.

\begin{Remark}\rm
When ${\rm spt}(\mu_V)$ 
is smooth, a direct computation \cite{simon} shows that 
condition \eqref{eqsta} implies that ${\rm spt}(\mu_V)$ 
is a time-like minimal submanifold of codimension $k$,
and $\mu_V$ coincides, up to a positive constant,
with the $(n+1-k)$-dimensional Minkowski area restricted to ${\rm spt}(\mu_V)$.
\end{Remark}

\subsection{The stress-energy tensor}\label{secstress}
We let 
\begin{eqnarray*}
T^{\alpha \beta}_\eps(\uu) &:=& - c_k(\eps) \tensoremetrico^{\alpha \gamma} 
\partial_{x^\gamma} \uu\,\cdot\,  
\tensoremetrico^{\beta \delta} \partial_{x^\delta} \uu 
\\
&& + \ell_\eps(\uu)\,\tensoremetrico^{\alpha \beta} 
\end{eqnarray*}
be the contravariant components of the symmetric
stress-energy tensor,
where $\cdot$ is the euclidean scalar product in $\R^k$.
Notice that 
\begin{equation}\label{Tepse}
|T^{\alpha \beta}_\eps(\uu)|\le e_\eps(\uu),
\end{equation}
for any $\alpha,\beta \in \{0,\dots,n\}$.
A direct computation shows that a solution $\uu_\eps$ of \eqref{eqhyp} 
satisfies
\begin{equation}\label{eqdiv}
\partial_{x^\beta} T^{\alpha \beta}_\eps(\uu_\eps) = 0.
\end{equation}
As a consequence, for every vector field $X \in C^1_c(\R^{1+n})$
we have
\begin{equation}\label{eqtest}
\int_{\R^{1+n}} \eta\, T^{\alpha \beta}_\eps 
(\uu_\eps)\,\partial_{x^\beta} X~dtdx = 0.
\end{equation}
Since  $\vert T^{\alpha \beta}_\eps(\uu_\eps)\vert$ 
is bounded by $e_\eps(\uu_\eps)$, 
 for any $\alpha, \beta \in \{0,\dots,n\}$
there exists a measure $T^{\alpha \beta}$
such that 
\begin{equation}\label{weakT}
T^{\alpha \beta}_\eps(\uu_\eps)dtdx\rightharpoonup T^{\alpha \beta}
\end{equation}
as $\eps \to 0^+$.
We denote by $T$ the measure-valued tensor
with components
$T^{\alpha\beta}$. Note that 
$T^{\alpha\beta}
= T^{\beta\alpha}$ and 
${\rm spt}(T)=\Gamma$. 

{}From \eqref{Tepse} it follows that $T^{\alpha\beta}$ on the right
hand side of 
\eqref{weakT} is absolutely
continuous with respect to $e$, 
and therefore is also absolutely continuous with respect 
to $\ell$, thanks to (A1). 
We denote by $\wT^{\alpha\beta}$ the density 
of the measure $T^{\alpha\beta}$ with respect to 
the measure $\ell$, i.e.,
\begin{equation}\label{deftildeT}
\widetilde T^{\alpha\beta} := \frac{dT^{\alpha\beta}}{d\ell},
\end{equation}
and by $\widetilde T$ the tensor
with components $\widetilde T^{\alpha\beta}$. 
 
In addition to (A2),
we shall also assume that 
\begin{itemize}
\item[(A3)] for $\mathcal H^{n+1-k}$-almost
every $x \in \Gamma$,\\ the tensor
${\rm Id}-\eta\widetilde T(x)$
is space-like. 
\end{itemize}
Recalling the expression of 
$T_\eps(\uu_\eps) - \ell_\eps(\uu_\eps)\tensoremetrico^{-1}$, 
we point out that (A3) is reminiscent to require 
that the tensor $\overline \nabla \uu_\eps$, suitably normalized,  
becomes space-like near $\Gamma$
as $\eps \to 0^+$, and that $\Gamma$ is time-like 
$\mathcal H^{n+1-k}$-almost everywhere. 
In particular, if $k=1$ and 
$\Gamma$ is smooth, the tensor ${\rm Id}-\eta\widetilde T$ 
turns out to be equal to 
$\eta^{-1}\nu_m\otimes \nu_m$, so that
$\eta \widetilde T$ is a lorentzian orthogonal projection, and
(A3) is equivalent to require that 
the normal vector $\nu_m$ to $\Gamma$ is space-like $\mathcal H^{n}$-almost everywhere.
This is for instance consistent with the explicit solution 
corresponding to a singular pulsating sphere considered in \cite{ViSh:94}.

\subsection{Main result}\label{secmain}

We are now in a position to prove 
the main result of the paper.

\begin{Theorem}\label{teomain}
Assume that the initial data $\uu_\eps(0,x)$ satisfy \eqref{ipoini}.
Let $\ell$, $w$ and $\widetilde T$ be defined 
as in \eqref{deflw}, \eqref{deflwbis} and
\eqref{deftildeT} respectively.
Under the assumptions (A1)-(A3), the following two statements hold:
\begin{itemize}
\item[(i)] Let $k=1$, and 
assume further 
\begin{equation*}
{\rm (A4)}\qquad \qquad \frac{dw}{dl}=\frac{1}{2}. 
\end{equation*}
Then 
$(\ell, \delta_{ \tensoremetrico \widetilde T})$
 is a stationary lorentzian rectifiable varifold
of codimension one.
\item[(ii)] Let $k=2$. Then 
$(\ell, \delta_{ \tensoremetrico \widetilde T})$
 is a stationary lorentzian rectifiable varifold
of codimension two.
\end{itemize}
\end{Theorem} 

As a consequence of Theorem \ref{teomain}, the set $\Gamma$ 
defined in \eqref{defGamma} is a rectifiable set 
of dimension $n+1-k$, and 
the tangent space to $\Gamma$ is time-like for $\mathcal H^{n}$-almost everywhere,
by assumption (A3). Moreover, in the regions where it is smooth, 
$\Gamma$ is a time-like minimal submanifold of codimension $k$,
and $\ell$ coincides, up to a constant, with the $(n+1-k)$-Minkowski area restricted to $\Gamma$.

Assumption (A4) corresponds to the so-called {\it equipartition}. 
In the parabolic case and when $k=1$, the analog of (A4)
turns out to be automatically satisfied
 \cite{Il:93}, and in that framework this property
shows that 
$\int_{\Rn} \eps \vert \grad u_\eps(t,\cdot)\vert^2~dx$ and 
$\int_{\Rn} \frac{1}{\eps} W(u_\eps(t,\cdot))~dx$ equally
contribute in the limit $\eps \to 0^+$. Still in the parabolic
case and when $k=2$, there is no equipartition,
and the contribution of 
$c_2(\eps) \int_{\Rn} \frac{1}{\eps^2} W(\uu_\eps(t,\cdot))~dx$ 
turns out to be negligible with respect to 
$c_2(\eps) 
\int_{\Rn} \vert \grad \uu_\eps(t,\cdot)\vert^2~dx$. This has an analog
in our hyperbolic case (see formula \eqref{noequipartition} below). 

\begin{proof} 
Passing to the limit in the linear condition 
\eqref{eqtest} we have
\begin{equation}\label{genvar} 
\int_{\R^{1+n}} \partial_{x^\beta} X\, d\, \eta T^{\alpha \beta} = 0,
\end{equation} 
for any $\alpha \in \{0,\dots,n\}$.
Note that \eqref{genvar} is (the generic component of) 
the stationarity condition 
for  the lorentzian varifold $(\ell, \delta_{\tensoremetrico 
\widetilde T})$.
Therefore, it is enough to prove that 
$(\ell, \delta_{\tensoremetrico 
\widetilde T})$ is a rectifiable lorentzian varifold,
i.e. for $\mathcal H^{n+1-k}$-almost every $x \in \Gamma$
the matrix
$\tensoremetrico \wT(x)$ is the lorentzian orthogonal 
projection onto the tangent space to $\Gamma$ at $x$.

By a rescaling argument around 
$\H^{n+1-k}$-almost every point $x\in\Gamma$ as in 
\cite[Eq. (3.6)]{AmSo:98}, from \eqref{genvar}
we obtain
\begin{equation}\label{eqphi}
\eta \wT(x) \int_{\R^{1+n}} \overline \nabla\phi \, d\nu =0,
\end{equation}
for all test functions $\phi$ supported in the euclidean unit
ball of $\R^{1+n}$, and for all $\nu$ in the tangent 
space to $\ell$ at $x$. 
As in \cite[Lemma 3.9]{AmSo:98}, from \eqref{eqphi} it follows that
for $\H^{n+1-k}$-almost every $x\in \Gamma$ 
\begin{equation}\label{atleast}
\!\!\!\!{\rm at~ least~} k {\rm ~eigenvalues~ of~} \eta \wT(x)
{\rm ~ are~ zero}.
\end{equation}
 These eigenvalues
correspond to the directions in the normal space to $\Gamma$ at $x$.

{}From the equalities 
\begin{equation*}
{\rm tr}(\partial_{x^\alpha} 
\uu_\eps \cdot \tensoremetrico^{\beta \delta} \partial_{x^\delta} \uu_\eps) 
=
- \vert{\uu_\eps}_t\vert^2 + \vert \nabla \uu_\eps\vert^2
\end{equation*} 
and 
\[
c_k(\eps)(\vert{\uu_\eps}_t\vert^2 - \vert \nabla \uu_\eps\vert^2) = 
2 \frac{c_k(\eps) W(\uu_\eps)}{\eps^2} - 2\ell_\eps(\uu_\eps),
\] 
we obtain 
\begin{equation}\label{eqtrace}
{\rm tr}(\eta T_\eps (\uu_\eps)) = 
2 \frac{c_k(\eps)W(\uu_\eps)}{\eps^2}
+ (n-1)\ell_\eps(\uu_\eps).
\end{equation}
Passing to the limit as $\eps\to 0^+$ we get 
\begin{equation}\label{eqtracelim}
{\rm tr}(\eta T) =  2w   +(n-1) \ell\,
\end{equation}
in the sense of measures. Considering the density with respect to $\ell$ 
we get
\begin{equation}\label{eqtracelimbis}
{\rm tr}(\eta \widetilde T) =   2\frac{dw}{d\ell}   +(n-1).
\end{equation}
Thanks to assumption (A3), for $\mathcal H^{n+1-k}$-almost
every $x \in \Gamma$ the tensor 
${\rm Id}-\eta \widetilde T(x)$, 
is space-like.  Therefore,
letting $\lambda_0,\lambda_1,\dots,\lambda_n$ 
be the eigenvalues of $\tensoremetrico \widetilde T(x)$,
 there exists a Lorentz transformation $L(x)$ such that 
\begin{eqnarray*}
&& L^{-1}(x)(\eta \widetilde T(x) - {\rm Id})L(x) 
\\
&=& L^{-1}(x)\,\eta \widetilde T(x)\,L(x) - {\rm Id} 
\\
&=& {\rm diag}(0,\lambda_1-1,\ldots,\lambda_n-1).
\end{eqnarray*}
In particular
$$
\lambda_0=1.
$$
Passing to the limit in the expression of $T_\eps(\uu_\eps) - \ell_\eps(\uu_\eps)
\tensoremetrico^{-1}$
 as $\eps \to 0^+$,
we get that 
$\widetilde T-\tensoremetrico^{-1} = \tensoremetrico^{-1} (\eta \widetilde T - 
{\rm Id})$ is 
negative semidefinite (in the euclidean sense), which implies
\begin{equation}\label{auto}
\lambda_i\le 1 \qquad 
\forall i \in \{1,\dots,n\}.
\end{equation}
{}From \eqref{auto} and \eqref{atleast} we then obtain
\begin{align}
{\rm tr}(\eta \widetilde T(x)) 
= \sum_{i=0}^n \lambda_i\le  n+1-k\,.
\label{antrace}
\end{align}
Note that
equality in \eqref{auto} holds if and only if $\eta \widetilde T(x)$
 is a lorentzian orthogonal projection  on a time-like
subspace of codimension $k$.
Consequently, 
our aim is now to prove equality in \eqref{antrace}.

{\it Case (i):} $k=1$.
Using (A4), \eqref{eqtracelimbis} becomes 
${\rm tr}(\eta \widetilde T(x)) =  n$.

{\it Case (ii):} $k=2$.
{}From \eqref{eqtracelimbis} and \eqref{antrace} it follows
$$
\frac{dw}{d\ell} \leq 0.
$$
Since $w$ is a positive measure, 
we deduce
\begin{equation}\label{noequipartition}
\frac{\d w}{\d \ell} =0.
\end{equation}
Therefore, \eqref{eqtracelimbis} becomes 
${\rm tr}(\eta \widetilde T(x)) =
 n-1$.
\end{proof}

In \cite{neu} it is shown by a formal asymptotic argument (made rigorous in \cite{Je:08})
that the thesis of Theorem \ref{teomain}
holds true when $k=1$, for well-prepared initial data and 
before the appearence of singularities. 
However, the ``small ripples'' example in \cite{neu}   
suggests that Theorem \ref{teomain} (i) 
 may not be true in general, without assuming (A4). 
In particular, differently from the parabolic case \cite{Il:93},
we expect that (A4) is not always satisfied for not well-prepared 
initial data.


\bibliographystyle{plain} 

\end{document}